\documentclass[twocolumn]{book}

\usepackage[pdftex]{graphicx}

\DeclareGraphicsExtensions{.pdf,.mps,.png,.jpg,.gif}

\usepackage{makeidx,universe}

\makeauthorindex
\BookTitle{~}
\CopyRight{~}

\begin{document} %******************************************

\pagenumbering{arabic}

\chapter{%
{\LARGE \sf
 Hadronic Contributions to $R$ and $g-2$ from \\
Initial-State-Radiation Data } \\
{\small 
To  appear in the proceedings for XXIX Physics in Collision, 
  International Symposium in Kobe, Japan, August 30 - September 2, 2009}
\\
{\normalsize \bf Denis Bernard}, from the BaBar collaboration \\
{\small \it \vspace{-.5\baselineskip}%***** Affiliations ***********
Laboratoire Leprince-Ringuet - Ecole polytechnique, CNRS/IN2P3, 91128 Palaiseau, France
} }

\AuthorContents{D.\ Bernard} 
\AuthorIndex{Bernard}{D.}

 \baselineskip=10pt %*******
 \parindent=10pt  %*******

\section*{Abstract} %******** Body of document starts.****************

I review the recent efforts to improve the precision of the
prediction of the anomalous moment of the muon, in particular of the
hadronic contribution of the vacuum polarization, which is the
contribution with the largest uncertainty.
Focus is given to the recent result for 
$e^{+}e^{-} \to \pi^{+}\pi^{-}$ by the BaBar collaboration, obtained
using events with radiation in the initial state.

\section{Introduction} %%%%%%%%%%%%%%%

\begin{figure}[t]%%%%%%%%%%%%%%%
 \begin{tabular}{cc}
 \begin{minipage}{.45\hsize}
  \begin{center}
   \includegraphics[width=.98\textwidth]{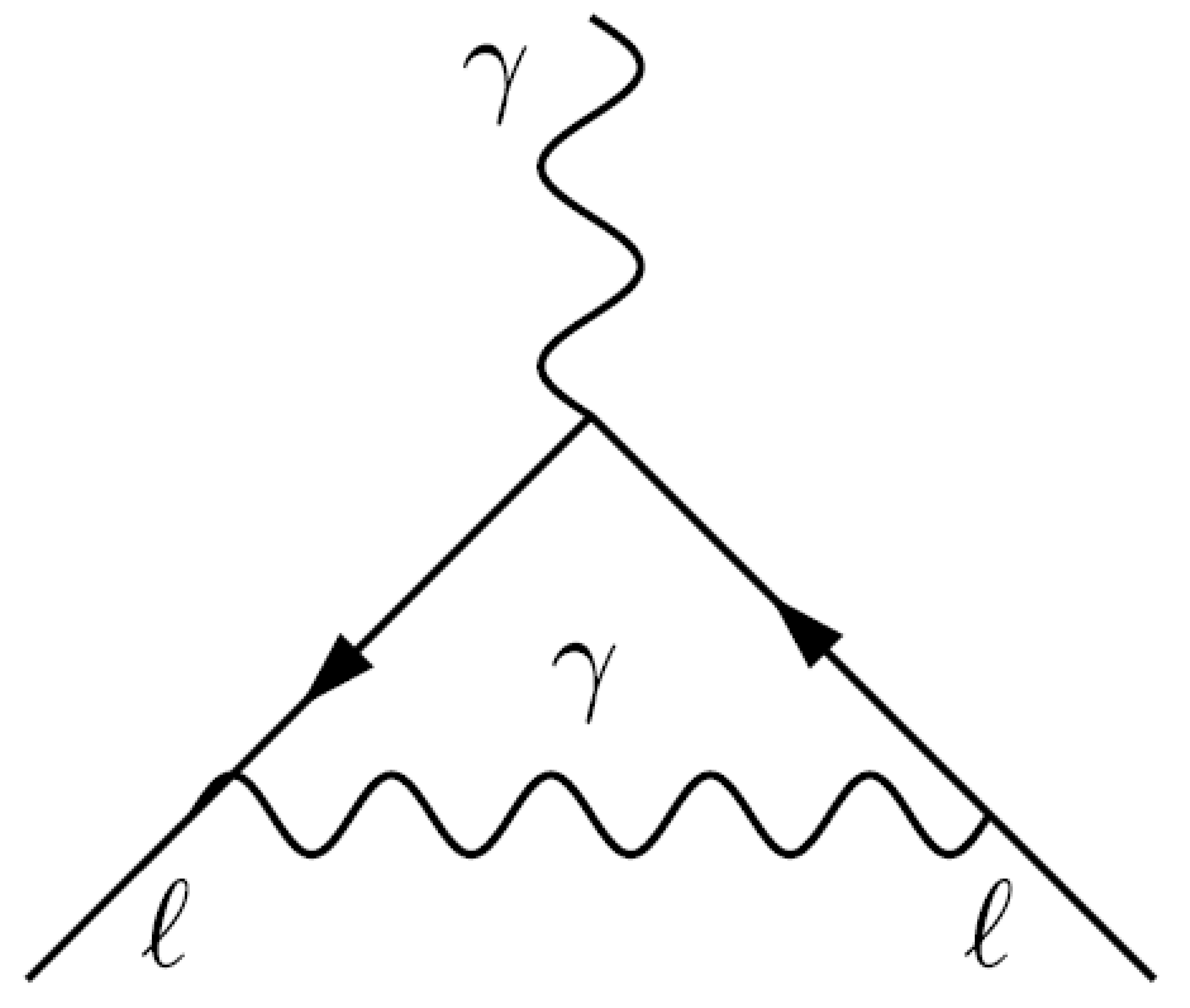}

   \caption{1st order contribution to $a$.}
  \label{fig:1st-order}
  \end{center}
 \end{minipage}

 \begin{minipage}{.45\hsize}
  \begin{center}
   \includegraphics[width=.98\textwidth]{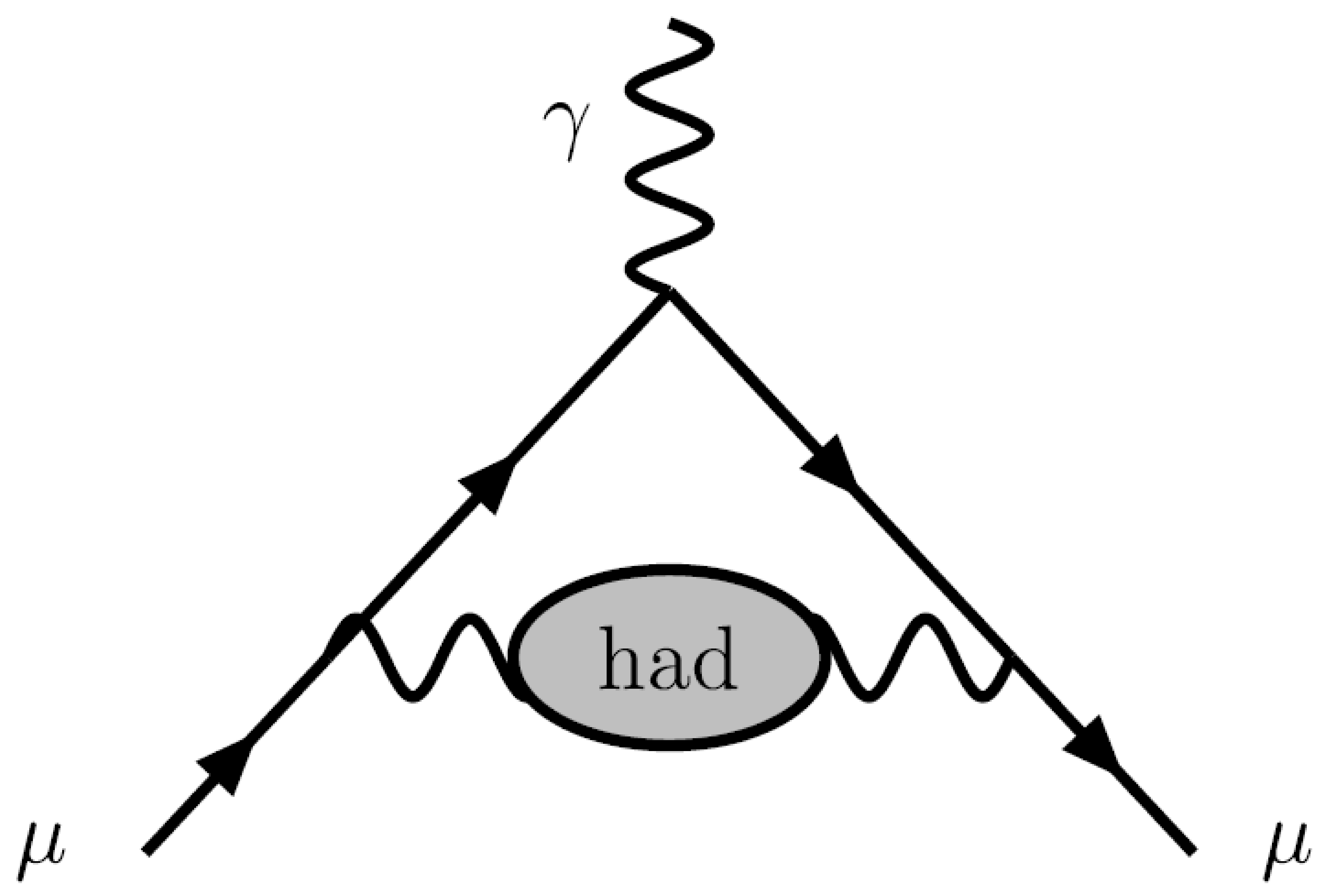}
   \caption{Lowest-order hadronic  VP diagram.}
  \label{fig:had-VP}
  \end{center}
 \end{minipage}
 \end{tabular}
% \end{figure}%%%%%%%%%%%%%%%

% \begin{figure}[t]%%%%%%%%%%%%%%%
 \begin{tabular}{cc}
 \begin{minipage}{.45\hsize}
  \begin{center}
   \includegraphics[width=.98\textwidth]{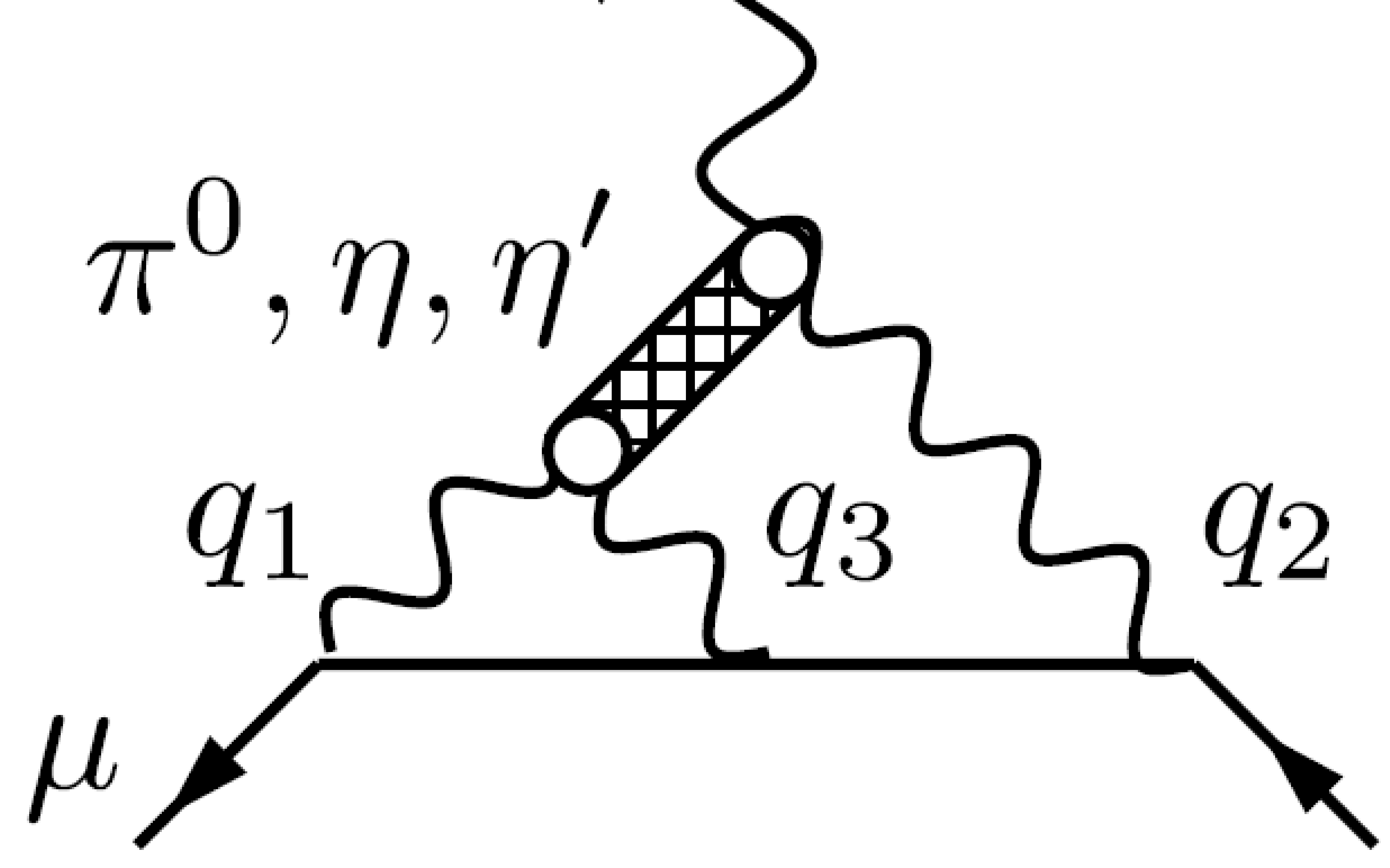}
   \caption{A light-by-light diagram.}
  \label{fig:light-by-light}
  \end{center}
 \end{minipage}
 \end{tabular}
\end{figure}

Elementary particles have a magnetic moment $\vec{\mu}$ proportional
to their spin $\vec{s}$, with $\vec{\mu} = (g e)/(2m) \vec{s}$. 
While pointlike Dirac particles would have $g=2$, i.e. an
``anomalous'' relative deviation of $a\equiv (g-2)/2 = 0$, Nafe {\em et
al.}  observed the first hints of a significant deviation from $a_e =
0$ more than 60 years ago \cite{Nafe:1947zz}.
The following year, Schwinger computed \cite{Schwinger:1948iu} the
first-order contribution to $a$, equal to $\alpha/(2\pi)$, the
diagram for which is shown in Fig. \ref{fig:1st-order}

The development of the quantum electro-dynamics (QED) followed, and
later of gauge theories in general, making these early works the very
basis of our present understanding of the elementary world.
Tremendous efforts have been devoted to improving the precision of the
theoretical prediction and of the direct measurement of $a$ since
then \cite{Jegerlehner:2009ry}.

More than 60 years later the situation is pretty exciting, with the
experimental and theoretical precision on the anomalous magnetic
moment of the muon $a_\mu$ both of the order of $6. \times 10^{-10}$,
and a discrepancy of $(29 \pm 9) \times 10^{-10}$ between them,
i.e. amounting to 3.2 $\sigma$, should Gaussian statistics be assumed
(Table \ref{tab:theory:exp}).

\begin{table}[h]
\small
\caption{Summary of the contribution to the theory prediction of the value of $a_\mu$, compared with the experimental measurement \cite{Jegerlehner:2009ry}.}
\begin{center} 
\begin{tabular}{lrrrr}
\hline
\hline
QED & 116 584 71.81 & $\pm$ 0.02 \\
Leading hadronic VP & 690.30 & $\pm$ 5.26 \\
Sub-leading hadronic VP & -10.03 & $\pm$ 0.11 \\
Hadronic light-by-light & 11.60 & $\pm$ 3.90 \\
Weak (incl. 2-loops) & 15.32 & $\pm$ 0.18 \\
\hline
Theory & 11659179.00 & $\pm$ 6.46 \\
Experiment \cite{Bennett:2006fi} & 11659208.00 & $\pm$ 6.30 \\
\hline
Exp $-$ theory & 29.00 & $\pm$ 9.03 \\
\hline
\hline
\end{tabular}
\end{center}
  \label{tab:theory:exp}
\end{table}

%%%%%%%%%%%%%%%

% \begin{itemize}
%\item 
The largest contribution to $a_{\mu}$ is by far from QED but its
contribution to the uncertainty is negligible.
% \item
In terms of uncertainty,  
the main contribution is from the hadronic component of the one-loop
vacuum polarization (VP, Fig.
\ref{fig:had-VP}) and, to a lesser extent, from the hadronic component
of the light-by-light processes (Fig. \ref{fig:light-by-light}).
% \end{itemize}

The photon propagator with VP is obtained from the bare propagator by
 replacing the electric charge $e$ by the energy dependant quantity:
\begin{eqnarray*}
e^2 \rightarrow e^2 / [1 + (\Pi'(k^2) - \Pi'(0))], 
\end{eqnarray*}

where $k$ is the photon 4-momentum.
At low energy, hadronic processes are not computable with the desired
precision. Instead the VP amplitude $\Pi'(k^2)$ is obtained from the
dispersion relation:
\begin{eqnarray*}
\Pi'(k^2) - \Pi'(0) = \displaystyle\frac{k^2}{\pi} \int_{0}^{\infty} \displaystyle\frac{Im \Pi'(s)}{s(s - k^2 - i\epsilon)} \mbox{d} s, 
\end{eqnarray*}

which in turn is related through the optical theorem:
\begin{eqnarray*}
Im \Pi'(s) = \alpha(s) R_{\mbox{had}}(s) /3, 
\end{eqnarray*}

to the ratio:
\begin{eqnarray*}
R_{\mbox{had}}(s) = \sigma_{\mbox{had}} \displaystyle\frac{3s}{4\pi \alpha(s)} = 
 \displaystyle\frac{\sigma_{e^{+}e^{-}\to hadrons}}{\sigma_{e^{+}e^{-}\to \mu^{+}\mu^{-}}}. 
\end{eqnarray*}

\begin{figure}[t]
  \begin{center}
    \includegraphics[width=0.95\linewidth]{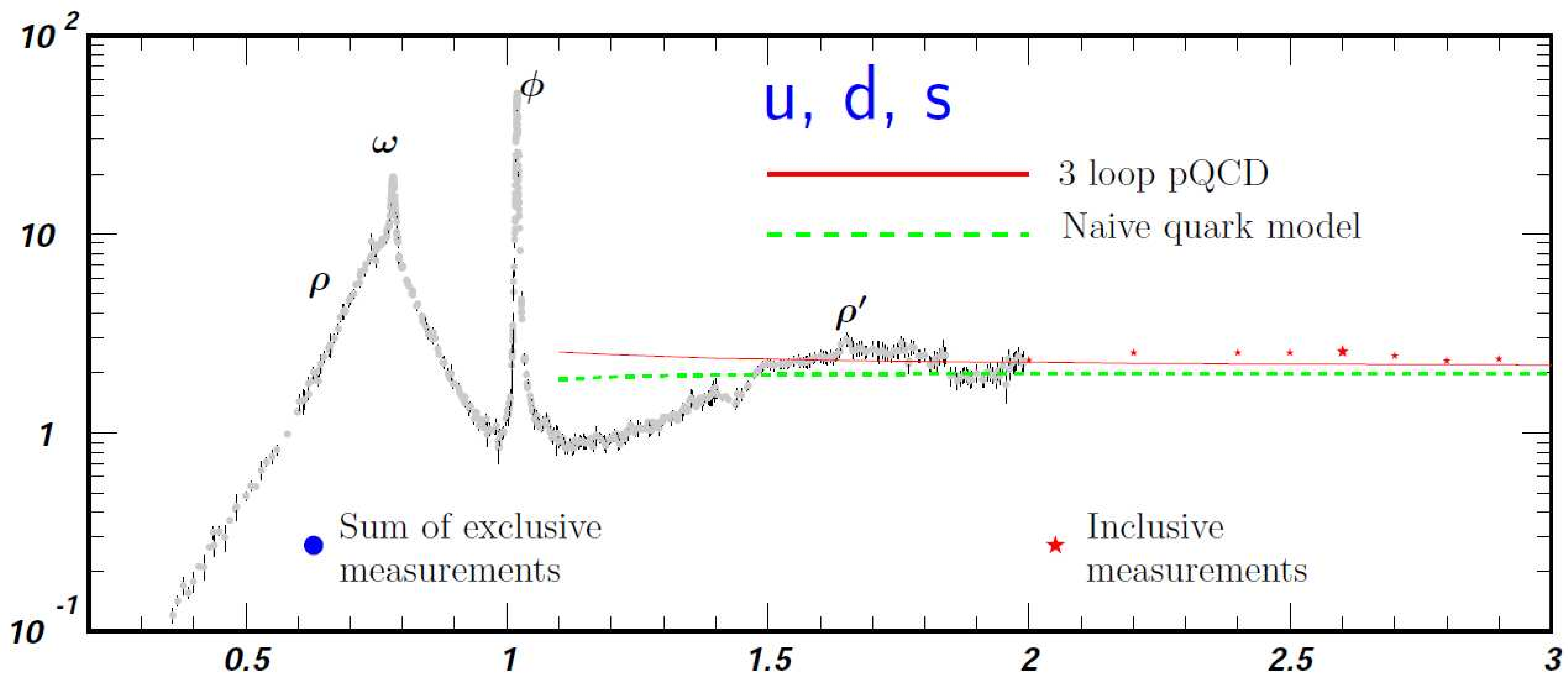}
%  \vspace{-1pc}
  \caption{$R_{\mbox{had}}$ as a function of $\sqrt{s}$ (GeV) \cite{Amsler:2008zzb}.}
    \label{fig:pdg} 
  \end{center}

  \begin{center}
    \includegraphics[width=0.95\linewidth]{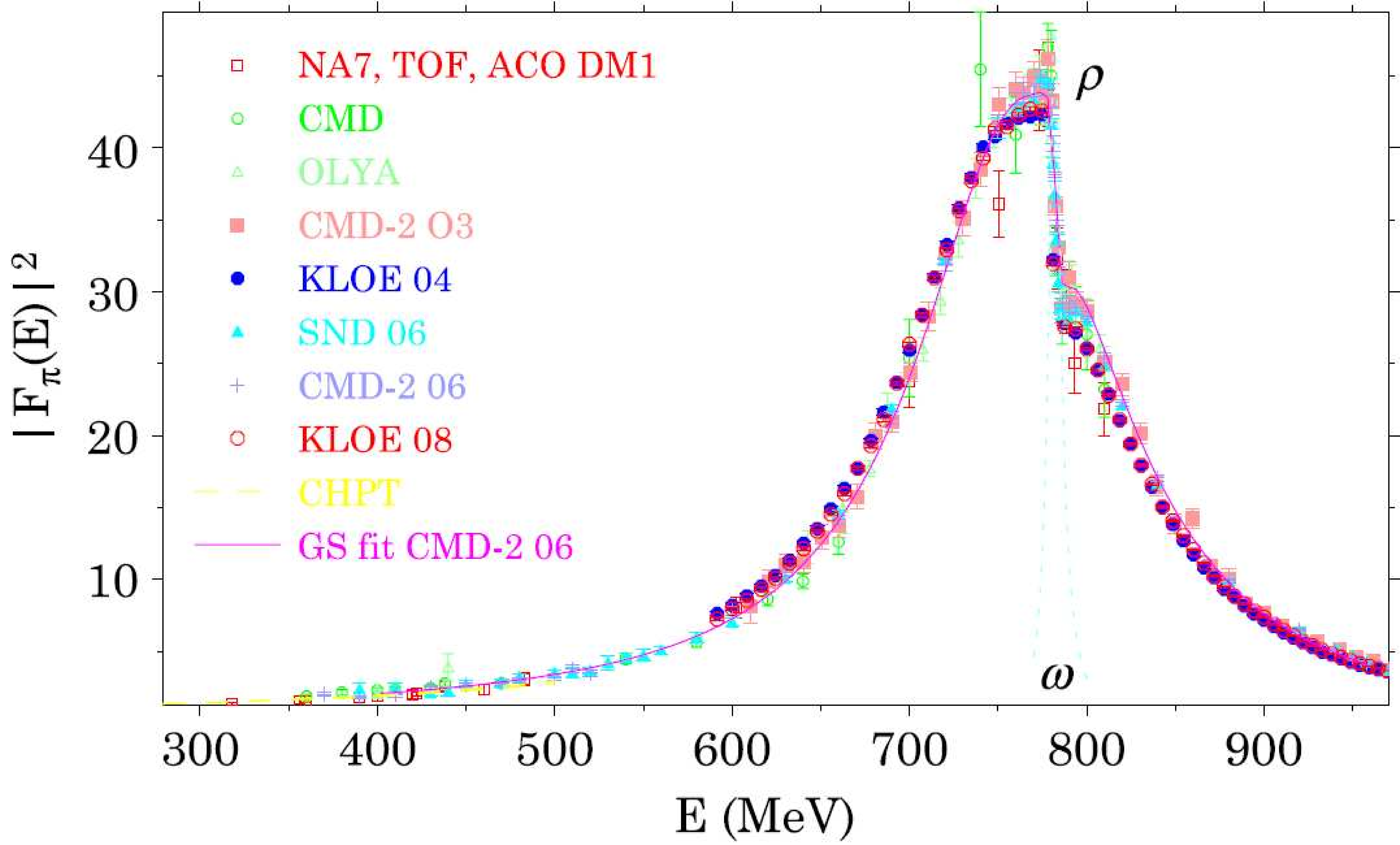}
%  \vspace{-1pc}
  \caption{Variation of the pion form factor squared with energy \cite{Jegerlehner:2009ry} (KLOE 04 is superseded by KLOE 08).}
    \label{fig:pipi:dir} 
  \end{center}
\end{figure}

Finally, the hadronic VP contribution is obtained from the
``dispersion integral'':
\begin{eqnarray*}
a_\mu^{\mbox{had}} = \left( \displaystyle\frac{ \alpha m_\mu} {3\pi} \right)^2
\int{ \displaystyle\frac{ R_{\mbox{had}}(s) \hat{K}(s)}{s^2} \mbox{d} s}, 
\end{eqnarray*}

where $\hat{K}(s)$ is an analytical function that takes values close to 1.
We note, from the $1/s^2$ variation of the integrand that the dominant
contribution comes from the low energy part of the integral.
A good experimental precision of the measurement of $R_{\mbox{had}}(s)$
at low energy is therefore welcome.
Fig. \ref{fig:pdg} shows a summary of the present measurements of
$R_{\mbox{had}}(s)$ \cite{Amsler:2008zzb}, where the presence of 
$J^{PC} = 1^{--}$ mesons can be seen.

The $\pi^{+}\pi^{-}$ channel has both the largest contribution and 
dominates the uncertainty, with 
$a_{\mu}^{\pi^{+}\pi^{-}}[2m_\pi, 1.8 \mbox{GeV}/c^2] = 
(504.6 \pm 3.1 (\mbox{exp}) \pm 0.9 (\mbox{rad}))\times 10^{-10}$,
compared to the full 
$a_\mu^{\mbox{had}} = (690.9 \pm 5.3)\times 10^{-10}$ from Table
\ref{tab:theory:exp}.
A summary of direct measurements in terms of the squared pion form
factor is shown in Fig. \ref{fig:pipi:dir}.
The 3.2 $\sigma$ discrepancy mentioned above is computed using this
$\pi\pi$ input.

\section{$\tau$ decay spectral functions}

The $I=1$ part of the $e^{+}e^{-} \to \pi^{+}\pi^{-}$ cross-section
can be estimated from the spectral function of $\tau$ decays to $\nu
\pi^{+}\pi^{0}$, under the hypothesis of conservation of the weak
vector current (CVC).
The method was pioneered by the ALEPH collaboration 
\cite{Barate:1997hv} and followed by OPAL 
\cite{Ackerstaff:1998yj}
and CLEO 
\cite{Anderson:1999ui}. 
Recently the Belle collaboration has performed an analysis using 
much larger statistics \cite{Fujikawa:2008ma}, obtaining a value
compatible with, and more precise than, the combination of all
previous results
\cite{Davier:2002dy}.

The $\tau$ method provides a high experimental precision, but 
extracting the contribution to $a_\mu$ depends on making a number
of isospin-breaking (IB) corrections. 
A recent update \cite{Davier:2009ag} of 
Ref. \cite{Davier:2002dy} lowers the correction by $\approx
7 \times 10^{-10}$, while the uncertainty on the correction is now
$1.5 \times 10^{-10}$. 

The branching fraction of the $\tau \to \nu \pi^{+}\pi^{0}$
decay also takes part in the calculation, with a 0.5 \% uncertainty.

\section{$e^{+}e^{-} \to \pi^{+}\pi^{-}$ using ISR method}

Initial-state radiation makes it possible to measure the cross-section
of the production of a final state $f$ in $e^{+}e^{-}$ collisions at a
squared energy $s$, over a wide range of energies, lower than
$\sqrt{s}$, through the radiation of a high energy photon by one of the
incoming electrons, after which the electrons collide at a squared
energy $s'$.

Recently the BaBar experiment has developped a systematic program of
measurements of cross-sections of $e^{+}e^{-}$ to hadrons at low energy,
using the ISR method \cite{babar:isr}.
The boost undergone by the final state $f$ provides an excellent
efficiency down to threshold.
In all studies by BaBar, the ISR photon is observed ($\gamma$-tag) and
its direction is compared to the direction predicted from the
direction of $f$, providing a powerful background noise rejection.
Most of these measurements are more precise than the
previously available results by about a factor of three.

\subsection{KLOE's result on $e^{+}e^{-} \to \pi^{+}\pi^{-}$} 

\begin{figure}[t]
  \begin{center}
    \includegraphics[height=10pc]{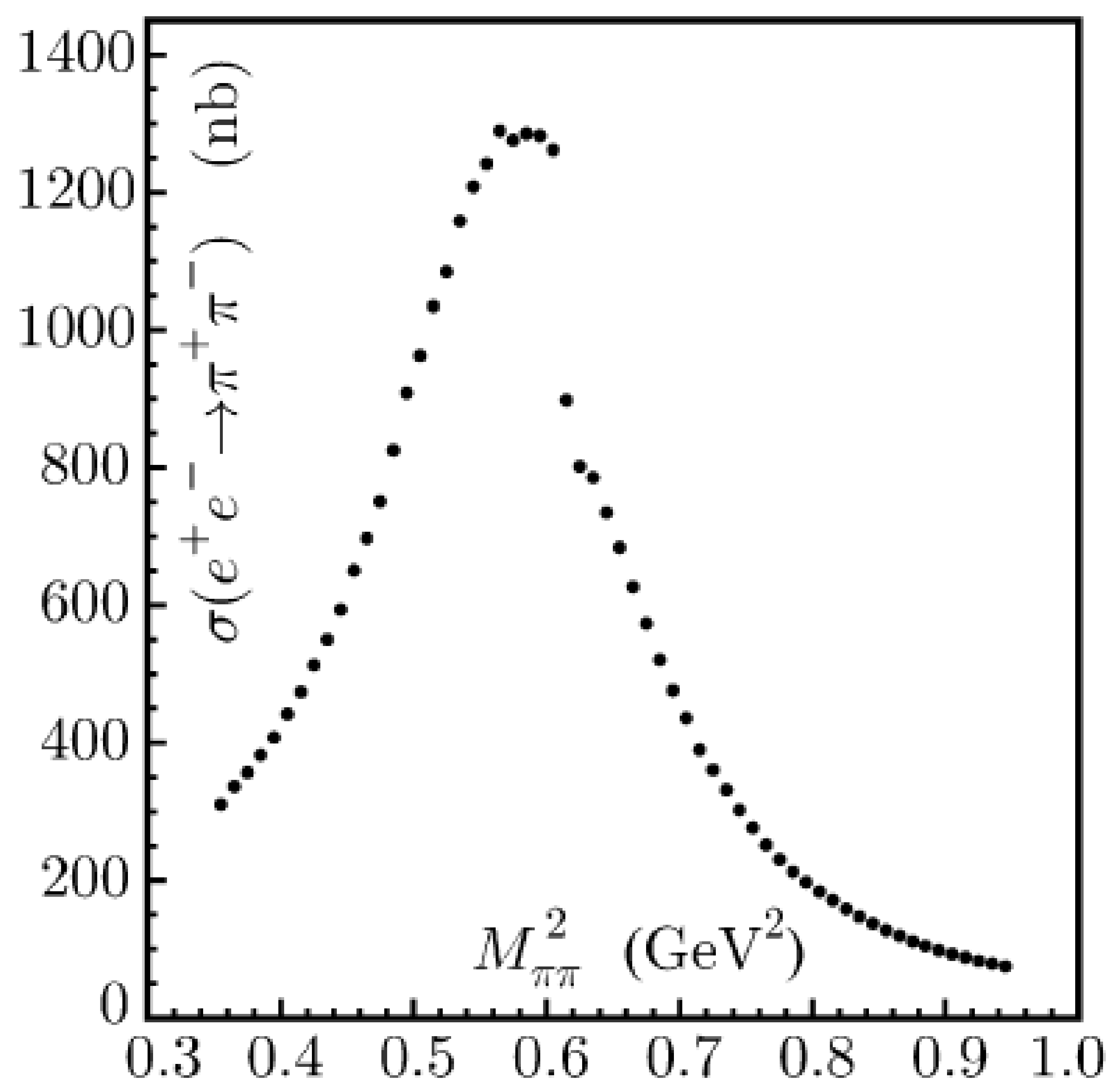}
%  \vspace{-1pc}
  \caption{$e^{+}e^{-} \to \pi^{+}\pi^{-}$ cross section measured by KLOE, using the ISR method \cite{:2008en}.}
    \label{fig:kloe} 
  \end{center}
\end{figure}

The KLOE experiment, when running on the $\phi$ resonance, has studied the
$e^{+}e^{-}$ annihiliations to $ \pi^{+}\pi^{-}$ with the ISR method
\cite{:2008en}, Fig. \ref{fig:kloe}. 
Here the ISR photon is not reconstructed: the request that the photon
direction be compatible with its having been emitted in the beam pipe
allows mitigation of the background to some extent, but the
systematical uncertainty on background subtraction is still a major
component of the total uncertainty.
The radiator function is provided from simulation, with systematics of
0.5\%, is the other major component.

The value of $a_{\mu}^{\pi^{+}\pi^{-}}$ obtained is compatible with
% and has similar precision as, 
the combination of previous results by CMD-2 \& SND over the mass
 range that they have in common of (630 - 958 MeV/$c^2$).

\subsection{BaBar's result on $e^{+}e^{-} \to \pi^{+}\pi^{-}$} 

BaBar uses a different approach: photon tagging is used, 
and the ISR luminosity is obtained from the muon channel, 
 $e^{+}e^{-} \to \mu^{+}\mu^{-} \gamma$ \cite{Aubert:2009fg}.
The systematics related to additional radiation is minimized in this
NLO measurement, i.e. radiation of one possible additional photon is
allowed, so that the final states actually reconstructed are
 $\pi^{+}\pi^{-} \gamma(\gamma)$ and 
 $\mu^{+}\mu^{-} \gamma(\gamma)$.
The ``bare'' ratio $R_{\mbox{had}}(s')$ mentioned above is obtained from the
experimentally measured $R_{exp}(s')$ after correction of final state
radiation (FSR) in
$e^{+}e^{-} \to \mu^{+}\mu^{-}$ and of additional FSR in ISR events 
 $e^{+}e^{-} \to \mu^{+}\mu^{-} \gamma $.

A number of important systematics cancel when measuring the $\pi/\mu$
ratio, such as those associated with the collider luminosity, the
efficiency of the reconstruction of the ISR photon, and the
understanding of additional ISR radiation.

The limiting factor is then the understanding of the possible
``double'' $\pi-\mu$, MC-data efficiency discrepancies. These are
studied in detail, with methods designed to disentangle correlations
as much as possible.
For example, inefficiency of the track-based trigger is studied using
events selected with a calorimetry-based trigger -- the small
correlation between both triggers being studied separately.
Likewise, $\mu$ and $\pi$ particle identification (PID) efficiency is
studied in good-quality, two-tracks ISR events, in which either one,
or both, tracks  meet the PID selection criteria.
Concerning tracking, a sizable degradation of the efficiency for
tracks overlapping in the detector was observed and studied in detail.

The systematics finally obtained are of the order of, or smaller than,
1 \% over the whole mass range studied; i.e., from threshold to 3
GeV/$c^2$.

\begin{figure}[t]
  \begin{center}
    \includegraphics[width=0.95\linewidth]{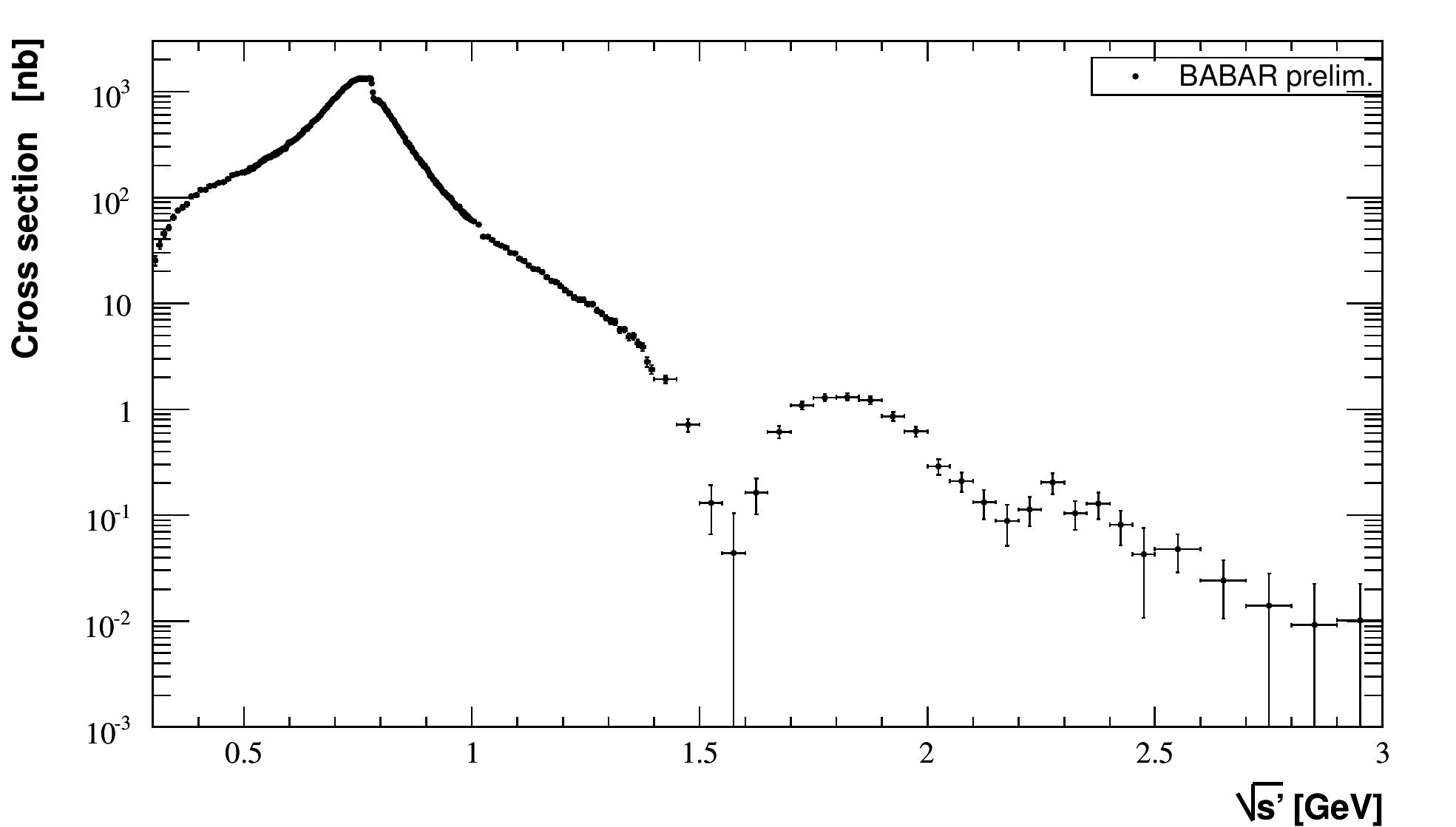}
%    \includegraphics[height=10pc]{BABAR_full.pdf}
%  \vspace{-1pc}
  \caption{Bare, unfolded 
$e^{+}e^{-} \to \pi^{+}\pi^{-}$ cross section measured by BaBar, using the ISR method \cite{Aubert:2009fg}.}
    \label{fig:babar:spectre} 
  \end{center}
\end{figure}

The $e^{+}e^{-} \to \pi^{+}\pi^{-}$ cross section measured by BaBar is
shown in Fig. \ref{fig:babar:spectre}. The sharp drop due to the
interference between the $\rho$ and the $\omega$, already present in
Fig.  \ref{fig:pipi:dir} is clearly visible.
The interference between the successive radial excitations of the
$\rho$ induces these dips in the cross-section.

The measured value of $a_{\mu}^{\pi^{+}\pi^{-}}[2m_\pi, 1.8\mbox{GeV}/c^2] =$
$ (514.1 \pm 2.2 \pm 3.1) \times 10^{-10}$
has a precision similar to the combination of all previous
$e^{+}e^{-}$-based results, but is larger by about 2.0 $\sigma$.

In addition to the measurement of the $\pi/\mu$ ratio, and extraction
of the $e^{+}e^{-} \to \pi^{+}\pi^{-}$ cross section, BaBar has
compared its $\mu^{+}\mu^{-}$ spectrum to the Monte Carlo prediction,
finding a good agreement within $0.4 \pm 1.1 \%$, dominated by the
collider luminosity uncertainty of $\pm 0.9 \%$.

\begin{figure}[t]
  \begin{center}
    \includegraphics[width=0.95\linewidth]{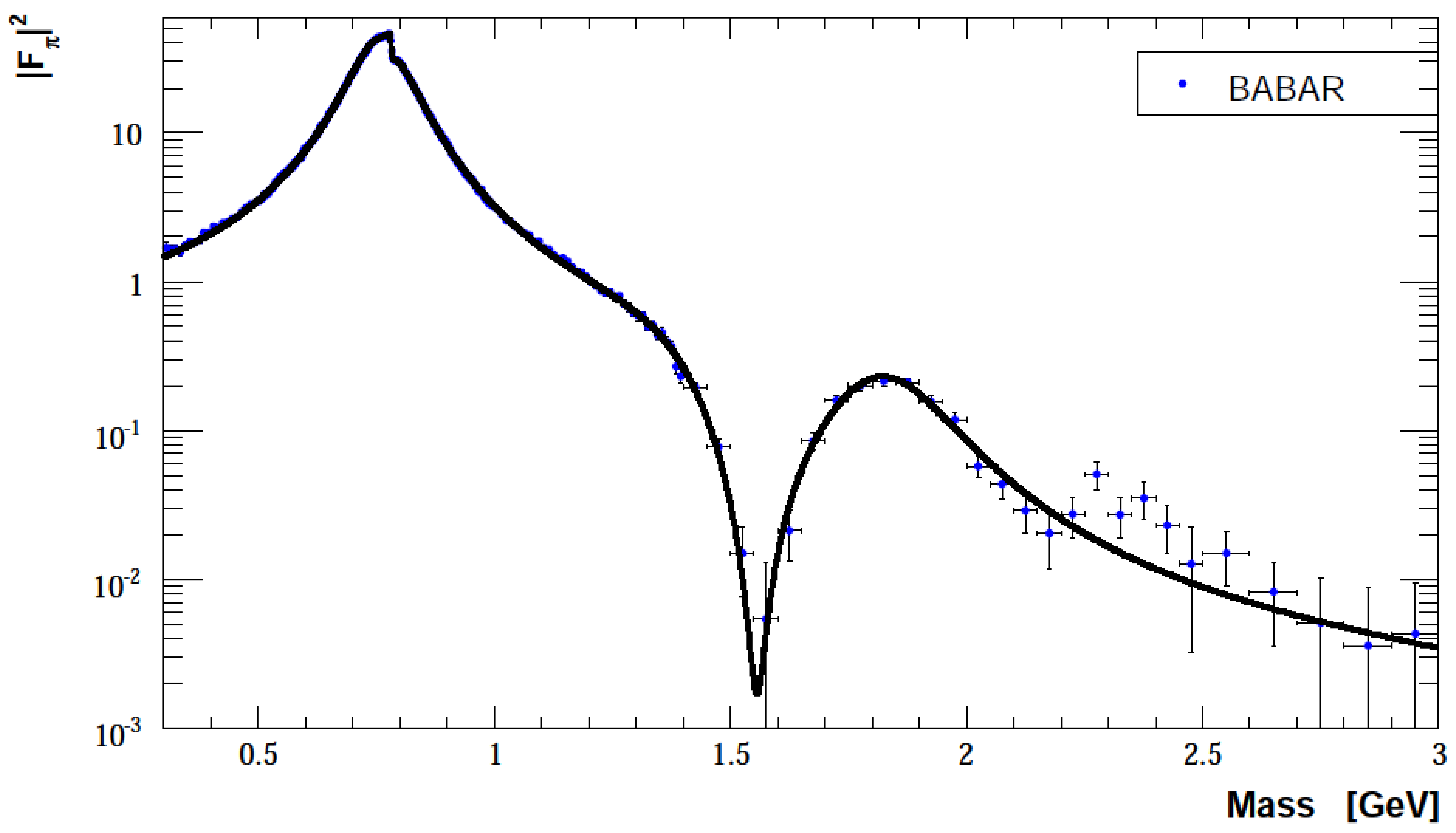}
%  \vspace{-1pc}
  \caption{VDM fit of the  squared pion form-factor by BaBar \cite{Aubert:2009fg}.}
    \label{fig:babar:vdm} 
  \end{center}
\end{figure}

\begin{figure}[t]
\begin{center}\small
\begin{tabular}{ccccccc} 
Belle $\tau$ & CMD2 $e^{+}e^{-}$ \\
\includegraphics[width=0.49\linewidth]{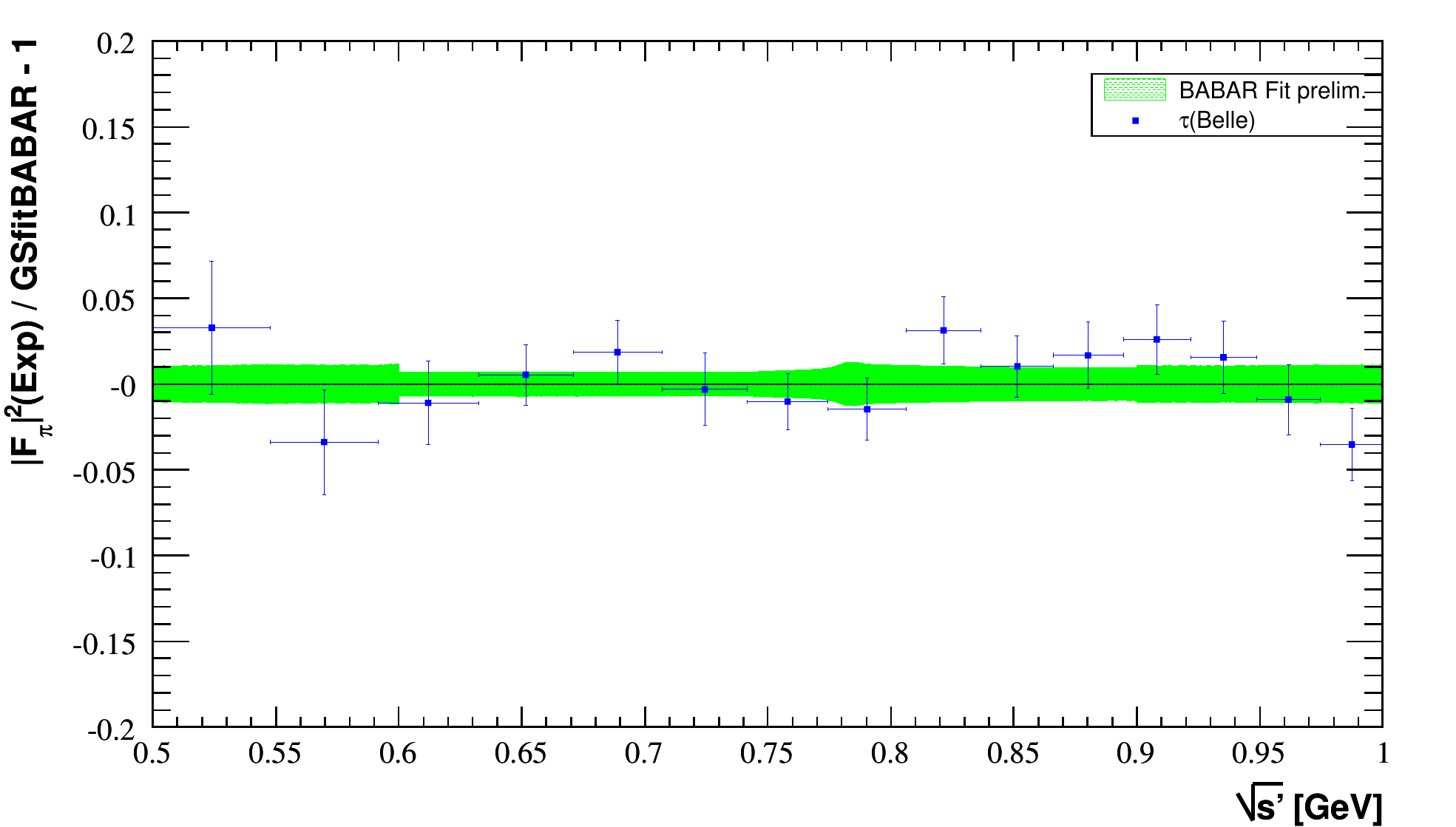}
&
\includegraphics[width=0.49\linewidth]{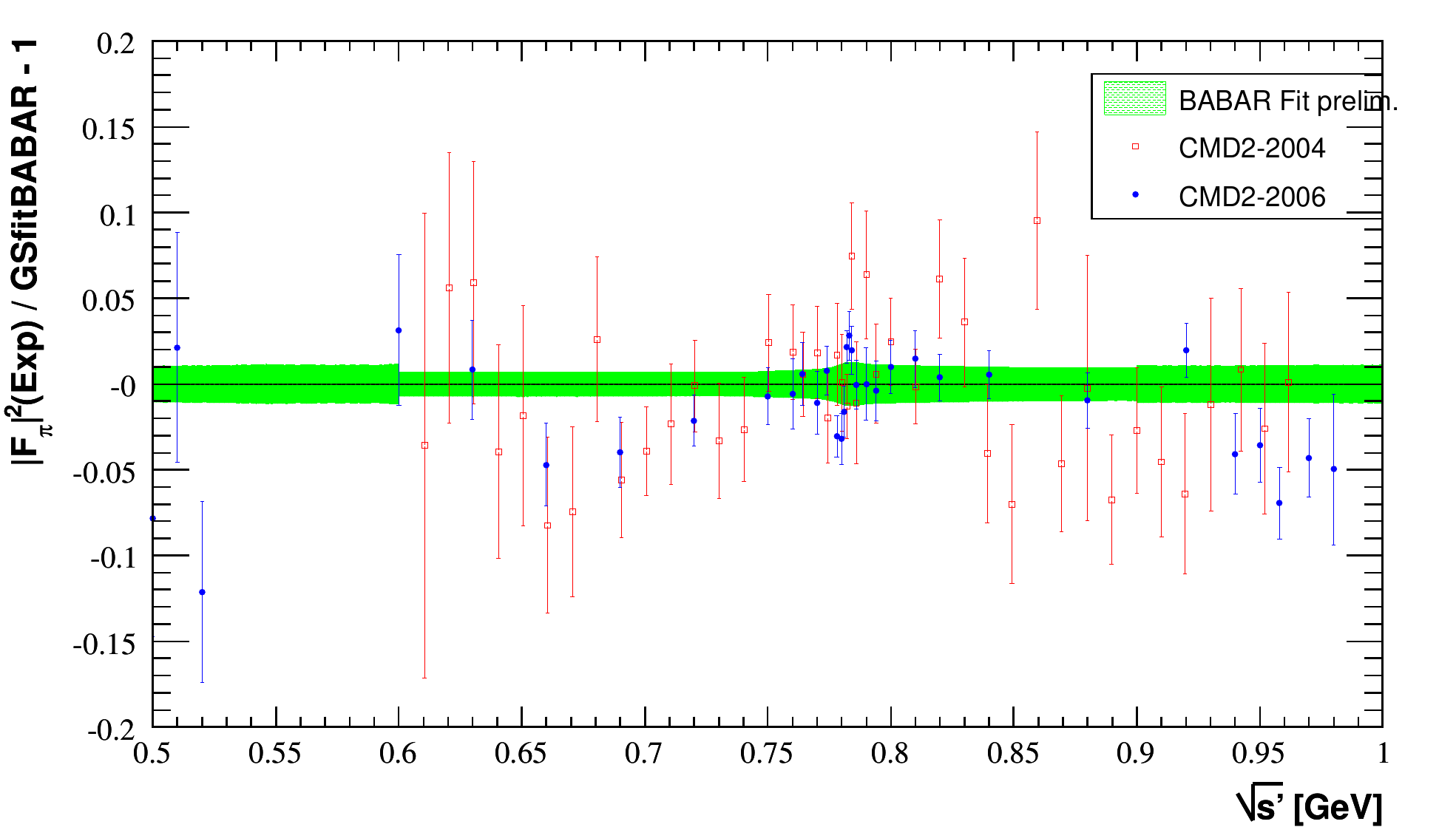}
\\
KLOE $e^{+}e^{-}$ ISR & SND $e^{+}e^{-}$ \\
\includegraphics[width=0.49\linewidth]{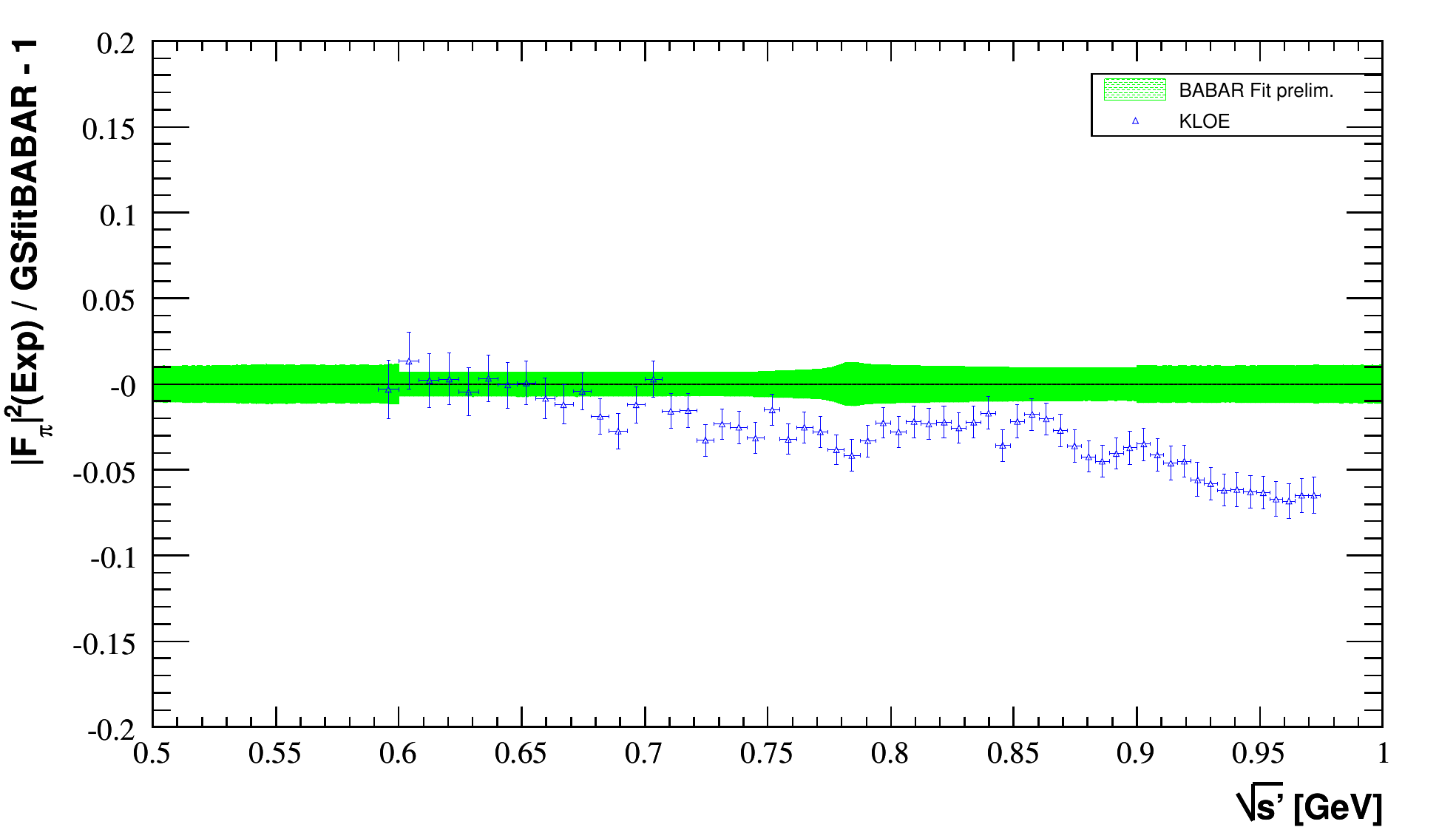}
&
\includegraphics[width=0.49\linewidth]{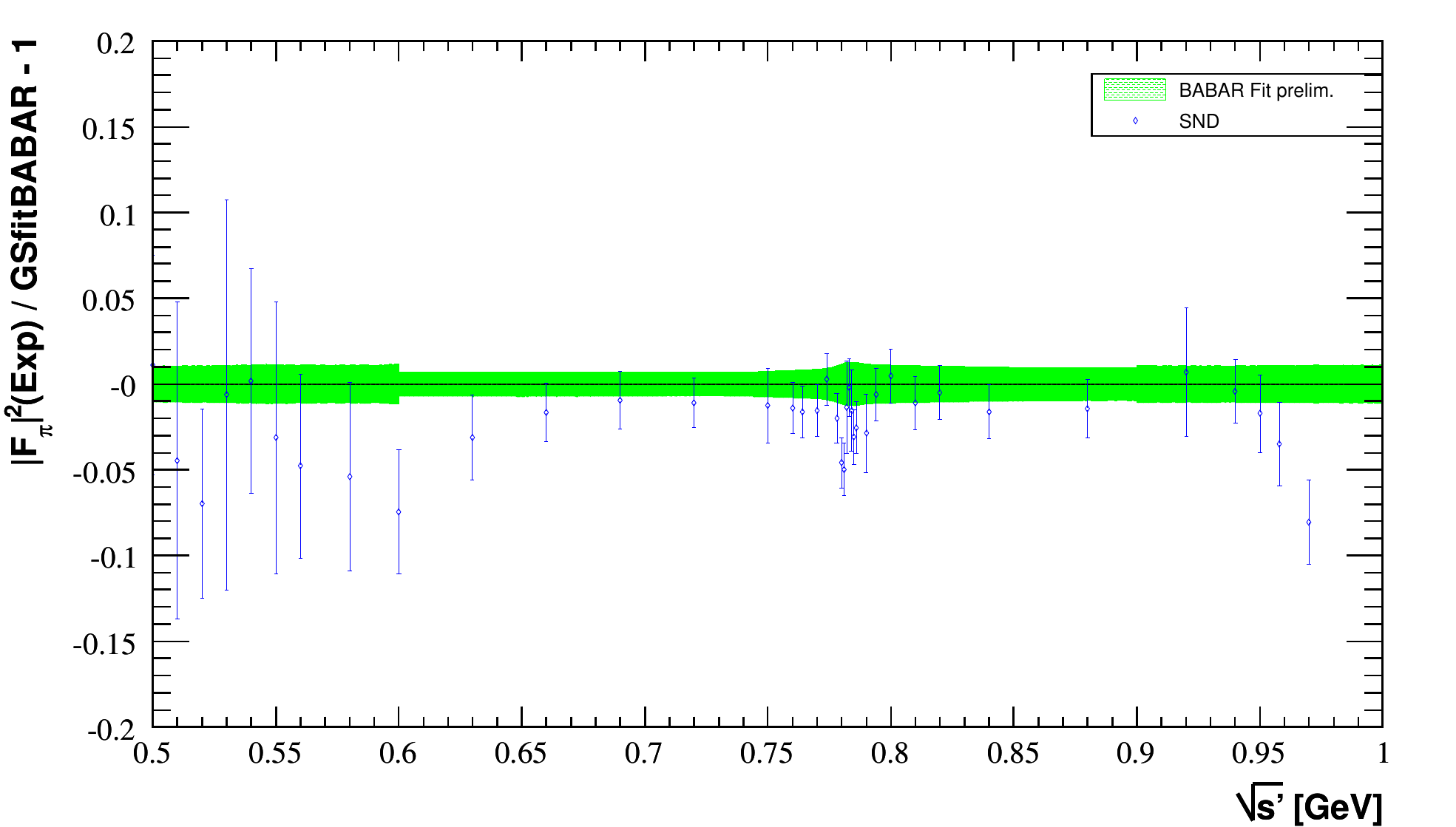}
\end{tabular}
%  \vspace{-1pc}
  \caption{Relative difference between the BaBar result, as fitted in Fig. \ref{fig:babar:vdm}, with that of previous experiments.}
    \label{fig:babar:comp} 
  \end{center}
\end{figure}

The distribution of the squared pion form-factor is fitted with a
vector-dominance model including the resonances $\rho, \rho', \rho'',
\omega$, with the $\rho$'s being described by the Gounaris-Sakurai
model.
The fit (Fig. \ref{fig:babar:vdm}) yields a good $\chi^2 /n_{df}$ of
$334 / 323$, and parameters compatible with the world-average values.
We can then use the fitted model to compare the result by BaBar with
 that of previous measurements (Fig. \ref{fig:babar:comp}).
The BaBar result is a bit larger than that obtained by CMD2\cite{Aulchenko:2006na} and SND 
\cite{Achasov:2005rg},
nicely compatible with the high-statistics $\tau$-based result by
Belle, but shows a clear disagreament with KLOE.

\section{$a_{\mu}$: the present situation}

The present situation in terms of $a_{\mu}$ is summarized in 
Fig. \ref{fig:babar:situation}:
\begin{figure}[t]
  \begin{center}
    \includegraphics[width=0.95\linewidth]{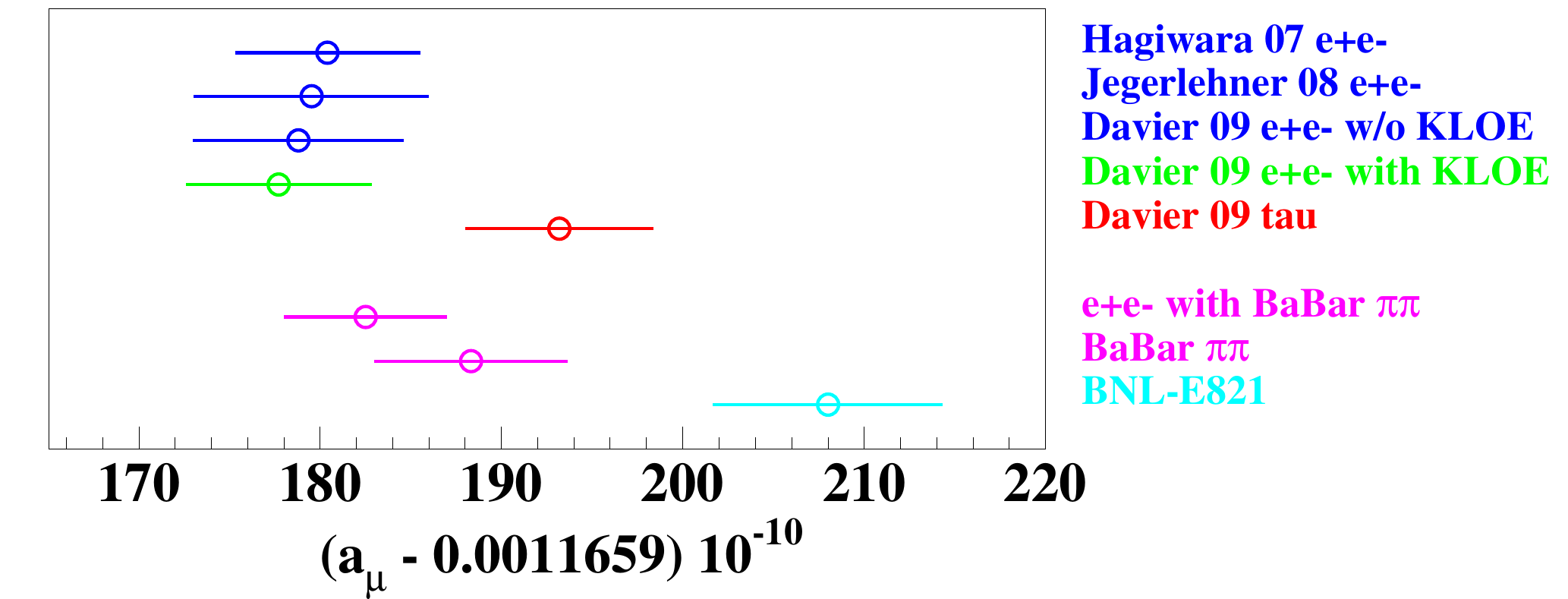}
%  \vspace{-1pc}
  \caption{$a_{\mu}$: the present situation.}
    \label{fig:babar:situation} 
  \end{center}
\end{figure}

\begin{itemize}
\item
The four upper points show that there is a general agreement between
 the various recent combinations of direct $e^{+}e^{-}$-based
 $\pi^{+}\pi^{-}$ measurement
\cite{Hagiwara:2006jt,Jegerlehner:2008zz,Davier:2009ag}.
\item
The large discrepancy between computations of $a_\mu$ based on these and the experimental measurement
by BNL-E821 \cite{Bennett:2006fi} is clear.
\item
The combination of $\tau$-based results, when corrected for
isospin-breaking effects using the most recent
calculation \cite{Davier:2009ag}, are also significantly lower than the
experimental measurement \cite{Bennett:2006fi}, by 1.8 $\sigma$.
\item
My computation of $a_{\mu}$ using the BaBar $\pi^{+}\pi^{-}$
measurement \cite{Aubert:2009fg} only is larger than the combination
of previous $e^{+}e^{-}$-based measurements, and compatible with the
$\tau$-based result, but still 2.4 $\sigma$ away from BNL-E821
\cite{Bennett:2006fi}.
\item 
The combination of all $e^{+}e^{-}$-based measurements, including the
recent one by BaBar, show an uncertainty that has decreased
significantly, and a central value that is larger.
However, the significance of the difference with respect to BNL-E821
 \cite{Bennett:2006fi} is barely changed, of the order of 3.3 $\sigma$.
\end{itemize}

After my talk, Changzheng Yuan  mentioned a soon-to-appear
preprint on ``Reevaluation of the hadronic contribution to the muon
magnetic anomaly using new cross section data from BABAR'' that is now
available \cite{Davier:2009zi}, and that obtains (more carefully)
similar results as that I have shown in Figure
\ref{fig:babar:situation}.

\section{What might take place in the next decade}

\subsection{$a_\mu$ measurement}

One single high-precision heavily-statistics-dominated experimental
measurement of $a_{\mu}$ \cite{Bennett:2006fi} is facing a prediction
in which the contribution with largest uncertainty has been 
confirmed, within reasonable significance, by a number of measurements
using various methods, each affected by its own systematics.

The obvious next step, before calling for new physics, is therefore to
check the measurement:
\begin{itemize}
\item A new
collaboration is planning to move the experimental apparatus from BNL
to FNAL, and perform a new measurement with  statistics increased by
a factor of 50, and reduced systematics, bringing the experimental
uncertainty down to 0.14 ppm, i.e., $1.6 \times 10^{-10}$
 \cite{Carey:2009zz}.
\item 
It would obviously be intensely desirable to cross-check such
a measurement using a completely different set-up. An alternative
 scheme is explored at J-PARC, with a micro-emittance muon beam
 inside a high-precision magnetic field, mono-magnet storage ``ring''
 \cite{Naohito:Saito}.
\end{itemize}

\subsection{Prediction}

On the prediction side, the main effort is understandably devoted to
the hadronic VP contribution.
\begin{itemize}
\item BaBar will complete its ISR program and
provide measurements of all possible hadronic final states in the low
energy range relevant to this discussion.
\item 
Belle may check BaBar's $\pi^{+}\pi^{-}$ measurement and BaBar may
check Belle's $\tau$ spectral functions.  KLOE is working on an
analysis with photon tagging too.
\item 
BES III will measure $R_{\mbox{had}}(s)$ in the range 2.0 -- 4.6 GeV,
something that will improve on $a_{\mu}$  only marginally, but will
also measure the $\tau \to \nu \pi^{+}\pi^{0}$ branching fraction with
improved precision \cite{Zhemchugov:2009zz}, an important ingredient
in the use of the $\tau$-based spectral functions.

The recent calculation of isospin-breaking corrections
\cite{Davier:2009ag} will be published and will doubtlessly be
cross-checked by other authors.
\item 
The collider at Novosibirsk has been upgraded to VEPP-2000
\cite{Shatunov:2008zz}, and the CMD 
\cite{Fedotovich:2006cq}
and SND experiments too.
\end{itemize}

Following the vacuum polarization, the next target in line for
improvement is the contribution of light-by-light scattering. Here
too work is in progress and there is hope to improve the precision,
both theoretically \cite{Prades:2008zz}, and using results of the
$\gamma\gamma$ programe at $DA\Phi NE-2$ \cite{Archilli:2008zz}

~

In total, there is good hope to bring both the prediction and
experimental uncertainties of $a_{\mu}$ at a very few $10^{-10}$.

\section{Acknowledgements}

I'd like to thank the organizers of PIC2009 for having invited me to
present this review and for having set the scene for such an
interesting gathering on the most recent results in particle physics
and high energy astronomy.

I am grateful to my collegues of the BaBar collaboration who helped me
to prepare this talk, with special thanks to Michel Davier.

%%%%%%%%%%

\begin{thebibliography}{99} \small 

\bibitem{Nafe:1947zz} 
J.~E.~Nafe, E.~B.~Nelson and I.~I.~Rabi, 
%``The  Hyperfine Structure Of Atomic Hydrogen And Deuterium,'' 
Phys.\ Rev.\ {\bf 71}, 914 (1947).

\bibitem{Schwinger:1948iu}
 J.~S.~Schwinger,
 %``On Quantum Electrodynamics And The Magnetic Moment Of The Electron,''
 Phys.\ Rev.\ {\bf 73}, 416 (1948).

\bibitem{Jegerlehner:2009ry}
 F.~Jegerlehner and A.~Nyffeler,
 %``The Muon g-2,''
 Phys.\ Rept.\ {\bf 477}, 1 (2009).

\bibitem{Bennett:2006fi}
 G.~W.~Bennett {\it et al.} [Muon G-2 Collaboration],
 %``Final report of the muon E821 anomalous magnetic moment measurement at BNL,''
 Phys.\ Rev.\ D {\bf 73}, 072003 (2006)
% [arXiv:hep-ex/0602035].

\bibitem{Amsler:2008zzb}
 C.~Amsler {\it et al.} [Particle Data Group],
 %``Review of particle physics,''
 Phys.\ Lett.\ B {\bf 667}, 1 (2008).

\bibitem{Barate:1997hv}
 R.~Barate {\it et al.} [ALEPH Collaboration],
 %``Measurement of the spectral functions of vector current hadronic tau decays,''
 Z.\ Phys.\ C {\bf 76}, 15 (1997).

\bibitem{Ackerstaff:1998yj}
 K.~Ackerstaff {\it et al.} [OPAL Collaboration],
 %``Measurement of the strong coupling constant alpha(s) and the vector and
 %axial-vector spectral functions in hadronic tau decays,''
 Eur.\ Phys.\ J.\ C {\bf 7}, 571 (1999)

\bibitem{Anderson:1999ui}
 S.~Anderson {\it et al.} [CLEO Collaboration],
 %``Hadronic structure in the decay tau- --> pi- pi0 nu/tau,''
 Phys.\ Rev.\ D {\bf 61}, 112002 (2000)

\bibitem{Fujikawa:2008ma}
 M.~Fujikawa {\it et al.} [Belle Collaboration],
 %``High-Statistics Study of the $\tau^- \to \pi^- \pi^0 \nu_\tau$ Decay,''
 Phys.\ Rev.\ D {\bf 78}, 072006 (2008)

\bibitem{Davier:2002dy}
 M.~Davier, S.~Eidelman, A.~Hoecker and Z.~Zhang,
 %``Confronting spectral functions from e+ e- annihilation and tau decays:
 %Consequences for the muon magnetic moment,''
 Eur.\ Phys.\ J.\ C {\bf 27}, 497 (2003)

\bibitem{Davier:2009ag}
 M.~Davier {\it et al.},
 %``The Discrepancy Between tau and e+e- Spectral Functions Revisited and the
 %Consequences for the Muon Magnetic Anomaly,''
 arXiv:0906.5443 [hep-ph], 
submitted to Eur. Phys. J. C. 

\bibitem{Hagiwara:2006jt}
  K.~Hagiwara, A.~D.~Martin, D.~Nomura and T.~Teubner,
  %``Improved predictions for g-2 of the muon and \alpha_{\rm QED}(M_Z^2),''
  Phys.\ Lett.\  B {\bf 649}, 173 (2007).
%  [arXiv:hep-ph/0611102].

\bibitem{Jegerlehner:2008zz}
  F.~Jegerlehner,
  %``Muon g - 2 update,''
  Nucl.\ Phys.\ Proc.\ Suppl.\  {\bf 181-182}, 26 (2008).


%\cite{babar:isr}
\bibitem{babar:isr}
 B.~Aubert {\it et al.} [The BABAR Collaboration], \\
 Phys.Rev.D77 (2008) 092002,
 Phys.Rev.D76 (2007) 092005,
 Phys.Rev.D76 (2007) 012008,
 Phys.Rev.D73 (2006) 052003,
 Phys.Rev.D73 (2006) 012005,
 Phys.Rev.D71 (2005) 052001,
 Phys.Rev.D70 (2004) 072004.

\bibitem{:2008en}
 F.~Ambrosino {\it et al.} [KLOE Collaboration],
 %``Measurement of $\sigma(e^+e^-\to\pi^+\pi^-\gamma(\gamma))$ and the dipion
 %contribution to the muon anomaly with the KLOE detector,''
 Phys.\ Lett.\ B {\bf 670}, 285 (2009)

\bibitem{Aubert:2009fg}
B.~Aubert  [The BABAR Collaboration],
  %``Precise measurement of the e+ e- to pi+ pi- (gamma) cross section with the
  %Initial State Radiation method at BABAR,''
  arXiv:0908.3589 [hep-ex], submitted to  Phys.Rev.Lett.

\bibitem{Davier:2009zi}
  M.~Davier, A.~Hoecker, B.~Malaescu, C.~Z.~Yuan and Z.~Zhang,
  %``Reevaluation of the hadronic contribution to the muon magnetic anomaly
  %using new e+e- -> pi+pi- cross section data from BABAR,''
  arXiv:0908.4300 [hep-ph].
%\cite{Carey:2009zz}




\bibitem{Aulchenko:2006na}
  V.~M.~Aulchenko {\it et al.}  [CMD-2 Collaboration],
  %``Measurement of the pion form factor in the energy range 1.04-GeV - 1.38-GeV with the CMD-2 detector,''
  JETP Lett.\  {\bf 82}, 743 (2005), 
%   [Pisma Zh.\ Eksp.\ Teor.\ Fiz.\  {\bf 82}, 841 (2005)]  [arXiv:hep-ex/0603021].
% \bibitem{Akhmetshin:2006wh}
  R.~R.~Akhmetshin {\it et al.},
  %``Measurement of the e+e- -> pi+pi- cross section with the CMD-2 detector in
  %the 370-520 MeV c.m. energy range,''
  JETP Lett.\  {\bf 84}, 413 (2006). 
%  [Pisma Zh.\ Eksp.\ Teor.\ Fiz.\  {\bf 84}, 491 (2006)]  [arXiv:hep-ex/0610016].




\bibitem{Achasov:2005rg}
  M.~N.~Achasov {\it et al.},
%``Study of the process e+ e- --> pi+ pi- in the energy region 400-MeV < s**(1/2) < 1000-MeV,''
  J.\ Exp.\ Theor.\ Phys.\  {\bf 101}, 1053 (2005)
%  [Zh.\ Eksp.\ Teor.\ Fiz.\  {\bf 101}, 1201 (2005)]  [arXiv:hep-ex/0506076].

\bibitem{Carey:2009zz}
  R.~M.~Carey {\it et al.},
 The New (g-2) Experiment, 
% A proposal to measure the muon anomalous magnetic moment to +-0.14 ppm precision,''
FERMILAB-PROPOSAL-0989

\bibitem{Naohito:Saito}
N. Saito, these proceedings.

\bibitem{Zhemchugov:2009zz}
  A.~Zhemchugov,
  %``Status Of Bes-Iii,''
  Nucl.\ Phys.\ Proc.\ Suppl.\  {\bf 189}, 353 (2009).

\bibitem{Shatunov:2008zz}
  Yu.~M.~Shatunov,
  %``VEPP-2000 collider commissioning,''
  Phys.\ Part.\ Nucl.\ Lett.\  {\bf 5}, 566 (2008).

\bibitem{Fedotovich:2006cq}
  G.~V.~Fedotovich  [CMD-3 Collaboration],
  %``Cmd-3 Detector For Vepp-2000,''
  Nucl.\ Phys.\ Proc.\ Suppl.\  {\bf 162}, 332 (2006).

\bibitem{Prades:2008zz}
  J.~Prades,
  %``Hadronic Light-by-Light Contribution to Muon g-2: Status and Prospects,''
  Nucl.\ Phys.\ Proc.\ Suppl.\  {\bf 181-182}, 15 (2008)

\bibitem{Archilli:2008zz}
  F.~Archilli, D.~Babusci, A.~D'Angelo, R.~Messi, D.~Moricciani and L.~Quintieri,
  %``Prospects for gamma gamma physics at DAFNE-2,''
  Nucl.\ Phys.\ Proc.\ Suppl.\  {\bf 181-182}, 248 (2008).


\end{thebibliography}
\end{document}